\documentclass[12pt]{article}

\begin{document}

\title{It is Time to Stop Teaching Frequentism to Non-statisticians}
\author{William M. Briggs \\300 E. 71st Apt. 3R, New York, NY 10021\\matt@wmbriggs.com}

\maketitle
\baselineskip=26pt

\begin{center}
{\bf William M. Briggs}  \\ \vskip .05in 300 E. 71st Apt. 3R, New York, NY 10021 \\ \textit{email:} matt@wmbriggs.com \\
\vskip .1in
\end{center}

\newpage

\vskip .1in\noindent \textsc{Abstract:}

We should cease teaching frequentist statistics to undergraduates and switch to Bayes.  Doing so will reduce the amount of confusion and over-certainty rife among users of statistics. 

\vskip .1in\noindent \textsc{Key words:} Bayes; Confidence intervals; Evidence; P-values; Teaching

\section{}
The time has come to cease teaching frequentist probability to non-statisticians.   No longer should undergraduates crease their brows to recall cookbook statistics.  Forever silent should graduate survey courses be on the subject of hypothesis tests.   P-values ought to be wrested from the grip of desperate scientists and dumped finally into the wastebasket of failed philosophical ideas.

Whatever beauty or truth is present in the mathematics of frequentist theory---and there is some---it ought to be left solely for initiates to pursue and to wonder over.   The occult mysteries of this curious doctrine, regardless how fascinating they are to the adept, should remain hidden from the outlander because we know {\it par exp\'{e}riences nombreuses at funestes} they are apt to misuse and misinterpret them. 

Just as Physics isn't Mathematics, but it has found tools from that subject useful, our field uses math but isn't math. The mathematics that arises in statistics can be interesting and rewarding; it is, anyway, the most rewarded pursuit for professional statisticians.  Oft forgotten, however, is that the fruit of our mathematical labors are not used as mathematics, of which few civilians love or care for, but to describe uncertainty, to answer vexing questions about evidence in real problems. Civilians aren't interested in the mathematical journey, but the practical destination.  

When supplicants approach us for wisdom about uncertainty, because of our ardent love of computation we have developed the unfortunate habit of insisting first on the memorization of mathematical incantations, such as figuring chi-squares (by hand),  or we require students look to up values in obscure tables in the backs of textbooks.   In pursuit of our love, we forget why civilians come to us. Even we statisticians can forget why the math is there.  Because of this, short shrift is paid to interpretation and to the limitations of the methods, to what it all means, which is all that civilians care about. 

Example number one: the confidence interval.  Its definition is so contrived and anti-intuitive that it's no wonder that students have difficulties ingesting it. To remind ourselves: all we can say about a given confidence interval, calculated from the data at hand, is that a (metaphysical) parameter either lies in the interval or that it does not. 

We are not allowed to say, though all cannot help but insist, that the parameter lies there with ninety-five percent probability.   And since the statement ``the parameter either lies in the interval or does not" is a tautology, the perceptive student is right to ask, ``Why bother?"  Why, indeed. Even trained statisticians, who should know better, err and treat the confidence interval as a credible interval. 

Which brings us to the second example: magical thinking in frequentist statistics.  The hunt for publishable p-values is nearly always fruitful.  If one cannot find a publishable p-value in one's data---with the freedom to pick and choose models and test statistics, to engage in ``sub-group" and sequential analysis, and so on---then one is being lazy.  P-values can and are used to prove anything and everything.  The sole limitation is the imagination of the researcher.  Fleeting exposure to a $72\times45$ pixel image of the American flag turns one into a Republican; walking through a door (an ``event segmentation") damages your memory; Keynesian theory is wrong; Keynesian theory is right; selenium causes cancer; selenium cures cancer.  This list could (and will) go on in perpetuity.  

This state of affairs is odd because frequentist theory tells us that the p-value is as silent as the tomb about the truth of the theory at hand.   Yet when a civilian cocks his ear towards a wee p-value, he hears music.  Angels sing.    To the civilian, the small p-value says that statistical {\it significance} has been found, and this in turn says that his hypothesis is not just probable, but true.  When a small p-value appears, it is if the result has been blessed.  Publication, and all that it happily entails, looms.

True a hypothesis may be, but since a scientific hypothesis speaks of tangible, real-world objects, with that which is observable, the p-value is not evidence whether the hypothesis is true or false. This is so {\it by design.}  The p-value is meant to be and is merely an indirect statement about the value of a mathematical creation assuming some nonexistent and unobservable parameters take a pre-specified value, a value which nobody believes these parameters can possibly take.  It used to be in logic that use of the Straw Man was seen as fallacious, yet the entire practice of frequentist statistics is devoted to this wispy device.

Cheating with and the misuse of p-values is rife. Misunderstanding is omnipresent. Every earnest exhortation---and these have been legion---to use these creations with caution has failed.  The mystical belief in their powers cannot be eradicated. Regardless of the purity of one's soul, the merest glance at a p-value less than the publishable limit seduces irrevocably. None can resist the mesmeric call of the p-value!  Best, then, to forbid their use and return them behind the veil.  

What should take the place of frequentism?  Equipped only with their common sense and ignorant of the fine distinctions of philosophy we statisticians spend years assimilating, civilians before they come to us think like Bayesians.  We do our best, by repetition and by rote, to beat this out of them.  But we have seen that they ill remember these lessons.  Further, the training is hateful to them, and once they are out of our sight they revert to their native thinking.

Bayesian statistics should, with the start of the next academic year, replace all undergraduate and casual  graduate frequentist statistics courses. Frequentism can remain for specialists (PhD students).  This wholesale change is the only chance we have of our students remembering what it is they are taught.  And when we make the switch, we had better cease emphasizing mathematical purity and focus on imbuing our charges with understanding the tools which they use.  Accentuate epistemology, deemphasize calculation. Highlight causes of over-certainty.

Since none but the devout want math, skip it or minimize it.  We will not lack for students: even without a saturation of mathematics, those who hear the call will still enter the fold.   Since civilians want to understand uncertainty, teach it.  This is our great forgotten purpose.  Since all are native Bayesians, face the inevitable.   Bayes certainly won't stop the flow of bad science but it will stem it---more so the faster we abandon our slavish and perplexing fascination with non-observable parameters.  Incidentally, Bayes allows a natural mechanism to do so: predictive statistics.

Besides the desirable decrease in over-certainty, the great Bayesian Switch will have many other benefits.  For one, the rescue of statistics courses taught by non-statisticians.  These erstwhile, yet unaffiliated instructors won't know the new material, you see, and will be forced to refer students to actual statisticians.  The pecuniary interest to statistics departments is obvious.  

Recovery of these courses is important because there are lot of folks out there who, because they once had a graduate survey course in regression, and have personally produced a p-value or two, feel they are versed sufficiently in probability to pass on their skills to fresh students.   But their rendering of the subtleties of the subject is often not unlike the playing of a worn cassette tape, which itself was copied from a bootleg.   This might even be acceptable, except that some who hear these concerts go on themselves to teach statistics.   This is an odd situation.  It is as if a cadre of mathematicians who had once read a greeting card and then wrote a love poem felt that they were qualified to teach English Literature and then began doing so---and issued credits for it, too---all with the hope that their students will go on to write novels.

Switching to Bayes will also be a boon to textbook publishers.  Out with the old and in with the new editions!  

But the largest contribution will be the dramatic reduction of people saying what they don't mean. Few or no civilians make sense when they discuss in Bayesian terms results from a frequentist analysis.   Civilians just can't remember that it is forbidden in frequentist theory to talk of the probability of a theory's or a hypothesis's truth.  They insist on translating the certainty they have in the value of some test statistic via the p-value to certainty that their hypotheses are true, despite that this is impossible to do so in frequentist theory. The result is that too many people are too certain of too many things.  
\end{document}